\documentclass[12pt]{article}
\usepackage{graphics}
\usepackage{epsfig}
\usepackage{latexsym}
\newcommand{\beq}{\begin{equation}}
\newcommand{\eeq}{\end{equation}}
\newcommand{\beqa}{\begin{eqnarray}}
\newcommand{\eeqa}{\end{eqnarray}}
\newcommand{\beqar}{\begin{eqnarray*}}
\newcommand{\eeqar}{\end{eqnarray*}}

\newcommand{\del}{\delta}

\newcommand{\ie}{{\it i.e.,}\ }
\newcommand{\labell}[1]{\label{#1}} 
\newcommand{\reef}[1]{(\ref{#1})}

{\bf }

\def\IR{{\hbox{{\rm I}\kern-.2em\hbox{\rm R}}}}
\begin{document}

\thispagestyle{empty}
\rightline{\small hep-th/0106038 \hfill McGill/01-11}
\vspace*{2cm}

\begin{center}
{ \LARGE The Entropy of 4D Black Holes and the Enhan\c{c}on  }\\[.25em]
\vspace*{1cm}

Neil R. Constable \footnote{E-mail: constabl@hep.physics.mcgill.ca}
\\
\vspace*{0.2cm}
{\it Department of Physics, McGill University}\\
{\it Montr\'eal, QC, H3A 2T8, Canada}\\
\vspace*{0.2cm}

\vspace{2cm} ABSTRACT
\end{center}
We consider the physics of enhan\c{c}ons as applied to four dimensional
black holes which are constructed by wrapping both D-branes and 
NS-branes on K3.
As was recently shown for the five dimensional
black holes, the enhan\c{c}on is 
crucial in maintaining consistency with the second law of thermodynamics.
This is true for both the D-brane {\it and} NS-brane sectors of these
black holes. In particular NS5-branes in both type IIA and IIB string theory 
are found to exhibit enhan\c{c}on physics when wrapped on a K3 manifold.

\vfill \setcounter{page}{0} \setcounter{footnote}{0}
\newpage

\section{Introduction}

One of the most fruitful applications of D-branes\cite{joe} 
in string theory has been the study of black holes. 
Most notably~\cite{stromv} D-brane
technology has been used to give a statistical mechanical description 
of the Bekenstein-Hawking entropy~\cite{bekhawk} of a large class of 
five~\cite{cm,horstr,jake1,jake2} and 
four~\cite{rrc,ms,hlm,4d} dimensional black holes
--- see refs.~\cite{juan,amanda,youm,bala,mathur,mirjamc} for extensive references and 
reviews. 

Essential to the embedding of charged black holes in string theory is the notion
of compactification. In general one begins by constructing a ten-dimensional 
solution of type II
supergravity which contains the desired charges. This solution is
 then dimensionally reduced on
an appropriate compact manifold so that the lower dimensional solution has an event
horizon with finite area and thus a non-trivial entropy. A fundamental element
of these constructions is that all of the stringy constituents are wrapped around 
some part of the compact manifold so that the final configuration appears as a point
like source in the lower dimensional space. 

It has been known for some time~\cite{curvcoup,ghm} that wrapping a Dp-brane around 
certain manifolds with non-trivial topology induces a charge
associated with a D(p-4)-brane onto the world volume of the original Dp-brane. 
In particular when the compactification manifold is K3 a unit of negative charge
is induced leading to a negative contribution to the total tension of the wrapped
brane~\cite{bbg}. When large numbers of wrapped branes are brought together so 
that there is a non-trivial back reaction on the geometry these facts have interesting
consequences. As was pointed out in ref.~\cite{jpp} these situations generically
give rise to regions of space time where wrapped Dp-brane probes of the 
geometry acquire
a negative tension. These regions also contain naked time-like 
singularities known as repulsons\cite{bern,kl,cvet}. 
It was proposed in ref.~\cite{jpp} that these regions 
are in fact unphysical and should be excised from
the space time altogether. Inside of the incision surface 
there is simply flat space. Dp-brane probes become 
tensionless on the (outer) boundary of the excised region and cannot 
proceed further into the space time.
Finally inside of the boundary locus there is an enhanced gauge symmetry; hence the 
name enhan\c{c}on. The consistency
of this proposal from the point of view of supergravity 
has recently been studied in ref.~\cite{jmpr}. 

In this paper we will consider a particular class of four dimensional 
charged black holes in type IIB string theory. 
The {\it asymptotic} charges correspond to  
$Q_1$ D1-branes embedded in the world volume of $Q_5$ D5-branes and  $Q_{F1}$ 
fundamental (F1-) strings inside $Q_{NS}$ NS5-branes. The reduction to four dimensions
is accomplished by compactifying on $K3\times T^2$ where both the D5-branes and
the NS5-branes are wrapped around K3 leaving a pair of effective strings wrapping
orthogonal cycles on the torus ----- see below.   
This black hole may be obtained from the 
more familiar D2/D6/NS5/P black hole considered in ref.~\cite{cvetyoum,ms,juan}
by acting with T-duality on the circle around which the momentum modes propagate.

There are several novel features of this particular black hole. First, up to the
interchange of cycles on the torus this configuration is self dual under
type IIB S-duality. Also, since this configuration involves orthogonal
NS5-branes and D5-branes it might be expected that for finite string coupling the
physics of $(p,q)$ webs~\cite{pqweb} will be important in understanding 
the dynamics of these black holes.
Further, due to the BPS property of this system, the two  
effective strings are independent and hence, as will be seen
below, there are two independent enhan\c{c}on effects. 

Since this black hole is obtained by wrapping D5-branes on K3 it is expected that 
the enhan\c{c}on mechanism will be relevant.  
Indeed it was recently
shown in ref.~\cite{5d} that for the case of five dimensional black holes involving
D5-branes wrapped on K3 the enhan\c{c}on mechanism plays a decisive role in 
ensuring that the string theory embedding of these black holes is consistent 
with the second law of thermodynamics. The same result is found here:
Enhan\c{c}on physics arises to prevent, in all cases, D5-branes from being absorbed
into the black hole when their absorption would serve to decrease the entropy. 

In addition to the D5-branes the 
NS5-branes are also wrapped around K3 and might therefore be expected to 
exhibit enhan\c{c}on physics. Recall that under 
S-duality D5-branes
are exchanged with NS5-branes and D1-branes are interchanged with F1-strings. 
Further, the enhanced gauge symmetry found inside of the enhan\c{c}on locus
should survive the journey to strong coupling and should therefore be present in any 
S-dual formulation.
As we shall see there is indeed  an enhan\c{c}on
effect for wrapped NS5-branes. In the context of four dimensional black holes 
the role of the
NS5-brane enhan\c{c}on is to prevent violations of 
the second law of thermodynamics. 

The remainder of this paper is organized as follows. Section 2 contains a discussion
of these four dimensional black holes and the enhan\c{c}on from the point of view
of supergravity. These results are then reproduced via probe calculations in 
section 3. Section 4 is a discussion of the implications of enhan\c{c}on physics
for the second law of thermodynamics as it pertains to the black holes considered 
here.
Motivated by the relevance of enhan\c{c}on physics for NS5-branes within black holes,
section 5 focuses on enhan\c{c}ons for both type IIA/B NS5-branes in isolation.
In section 6 conclusions are presented and some interesting open problems are 
discussed.

\section{Black Holes in Four Dimensions and the Enhan\c{c}on} 

The four dimensional black hole to be considered in this paper is constructed from
the orthogonal intersection of a D1/D5 bound state and an F1/NS5 bound state.
The D1-branes are aligned along the $x^4$ direction, the
D5-branes fill the directions
spanned by $(x^4,x^5,x^6,x^7,x^8)$, the NS5-branes fill $(x^5,x^6,x^7,x^8,x^9)$
and the F1-strings lie along $x^9$. The reduction to four dimensions is accomplished 
by wrapping $(x^5,x^6,x^7,x^8)$ on a K3 manifold leaving an effective D1/D5 string
in the $x^4$ direction and an effective F1/NS5 string in the $x^9$ direction. The 
effective strings are then each compactified on circles of radii $R_4$ and $R_9$ 
respectively, leaving a black hole in
$(t,x^1,x^2,x^3)$.

The ten dimensional Einstein frame metric for this D1/D5/F1/NS5 system is,
\beqa
ds^2&=&-h_1^{-3/4}h_5^{-1/4}h_{F1}^{-3/4}h_{NS}^{-1/4}dt^2 + h_1^{-3/4}h_5^{-1/4}
h_{F1}^{1/4}h_{NS}^{3/4}dx_4^2
+ h_1^{1/4}h_5^{3/4}h_{F1}^{-3/4}h_{NS}^{-1/4}dx_9^2 
\nonumber \\
&+& h_1^{1/4}h_5^{-1/4}h_{F1}^{1/4}h_{NS}^{-1/4}ds_{K3}^2 
+h_1^{1/4}h_5^{3/4}h_{F1}^{1/4}h_{NS}^{3/4}\left(dr^2+r^2d\Omega_2^2\right).
\labell{10dmet}
\eeqa
The dilaton, Ramond-Ramond and Kalb-Ramond potentials for this solution are,
\beqa
e^{2\phi}&=&\frac{h_{1}}{h_5}\frac{h_{NS}}{h_{F1}}
\nonumber \\
C_{(2)}&=&h_1^{-1} dt\wedge dx^4
\nonumber \\
C_{(6)}&=&h_5^{-1}dt\wedge dx^4\wedge\cdots\wedge dx^8 
\nonumber \\
B_{(2)}&=&h_{F1}^{-1}dt\wedge dx^9
\nonumber \\
B_{(6)}&=&h_{NS}^{-1}dt\wedge dx^5\wedge\cdots\wedge dx^9
\labell{bhdil}
\eeqa
where $B_{(6)}$ is the six form which couples electrically to the NS5-brane.
The harmonic functions in all of the above expressions are given by,
\beqa
h_{1}=1+\frac{c_{1}Q_{1}}{r} &\,\,\,\,\,\,\,&h_{5}=1+\frac{c_5Q_5}{r}
\nonumber \\
h_{F1}=1+\frac{c_{F1}Q_{F1}}{r}&\,\,\,\,\,\,\,&h_{NS}=1+\frac{c_{NS}Q_{NS}}{r}
\labell{4dharm}
\eeqa
and in four dimensions the constants $c_i$ are~\cite{juan},
\beq
c_{1}=\frac{gl_s^2}{2R_9}\frac{V^{\star}}{V}\,\,\,\,\,
c_{5}=\frac{gl_s^2}{2R_9} \,\,\,\,\,
c_{F1}=\frac{g^2l_s^2}{2R_4}\frac{V^{\star}}{V} \,\,\,\,\,
c_{NS}=\frac{l_s^2}{2R_4} 
\labell{4const}
\eeq
Notice that the local string frame volume of the K3 manifold is given by,
\beq
V(r)=\frac{h_{1}}{h_{5}}V
\labell{volK3}
\eeq
where $V$ is the asymptotic volume of the K3. For $c_1Q_1<c_5Q_5$ this volume is 
shrinking monotonically as we approach $r=0$. 
Specifically, we have $V(0)<V$ and so in accord with the
results of~\cite{jpp} we may expect that enhan\c{c}on physics
will be important at some finite value of $r$. Note that $V(r)$ is unaware of the
F1/NS5-brane component of the geometry and so we expect that when $V(r)=V^{\star}$
there will be an enhan\c{c}on locus beyond which individual D5-branes may not proceed
in a supersymmetric manner. Despite the fact that the local volume of K3 is independent
of the wrapped NS5-branes it will be found below that there is also a shell located 
at a finite radius which NS5-branes are forbidden from passing.  

As a final comment notice that the local value of the string coupling
at $r=0$ is given by
\beq
g^2e^{2\phi}=\frac{Q_1}{Q_5}\frac{Q_{NS}}{Q_{F1}}
\labell{coupling1}
\eeq
and is small provided that,
\beq 
\frac{Q_1}{Q_5} \ll \frac{Q_{F1}}{Q_{NS}}
\labell{weakcoup}
\eeq
in which case string loops may safely be neglected.

Upon reducing to four dimensions the surface at $r=0$ is 
an extremal horizon with vanishing surface gravity and non-vanishing area
which gives a Bekenstein-Hawking entropy,
\beq
S_{BH}=2\pi\sqrt{Q_1Q_5Q_{F1}Q_{NS}}
\labell{BH}
\eeq

Since this black hole is constructed from D5-branes wrapped on a K3 surface 
one expects that the enhan\c{c}on physics uncovered in ref.~\cite{5d} will be
relevant here. However,
as remarked in the introduction, S-duality suggests that wrapping an NS5-brane on
a K3 manifold will induce negative amounts of F1-string charge and 
therefore it seems that enhan\c{c}on physics will also be relevant
for this part of the system. This will be demonstrated later via an explicit 
probe calculation but for now it will simply be assumed and its consequences 
explored. 
With this in mind, recall that the integers $Q_1,Q_5,Q_{F1},Q_{NS}$ measure 
the asymptotic charges
associated with the constituents of the black hole. In order to facilitate
the following analysis it is convenient to introduce another set of  integers, 
$N_1,N_5,N_{F1},N_{NS}$, which
count the actual number of branes in the system. In particular $Q_1=N_1-N_5$ and
$Q_{F1}=N_{F1}-N_{NS}$ while $N_5=Q_5$ and $N_{NS}=Q_{NS}$.
In terms of these parameters the entropy in eqn.~\reef{BH} becomes,
\beq
S_{BH}=2\pi\sqrt{(N_1-N_5)(N_{F1}-N_{NS})N_{5}N_{NS}}.
\labell{BH2}
\eeq
Writing the entropy in this way suggests that this class of black holes is, 
for certain ranges of parameters, at odds
with the second law of thermodynamics. To see this assume that $N_1 = 2N_5$
and then consider slowly (\ie adiabatically) moving a single D5-brane from 
infinity into the black hole. 
It is not difficult to see that this will cause a decrease in the total entropy. 
Of course the same statement can be made about the 
F1/NS5 sector of this black hole.  See section 4 below for a complete discussion. 

For the case of five dimensional black holes constructed from D1-branes, D5-branes and
momentum modes it was shown in ref.~\cite{5d} that the resolution of this
paradox lies in the physics of the enhan\c{c}on mechanism~\cite{jpp}. In the
following it will be shown that the class of black holes under consideration
in fact require {\it two} enhan\c{c}ons to maintain consistency with the second law.

The physics pertaining to the D1/D5 sector of this black hole is already well
known~\cite{jpp,5d}. Since we expect that there exists a radius inside of which 
D5-branes cannot penetrate, the supergravity solution must be modified in the interior.
To model this, while at the same time remaining completely general, it is 
assumed that some number of the D-branes 
\ie  $\delta N_5$ D5-branes and $\delta N_1$ D1-branes, 
are evenly distributed over a two sphere located at a radius denoted by $r_c$.
The black hole in the interior therefore consists of  
$N_5^{\prime}=N_5-\del N_5$ D5-branes and $N_1^{\prime}=N_1-\del N_1$ D1-branes
corresponding to  asymptotic charges $Q_5^{\prime}=N_5^{\prime}$ and  
$Q_1^{\prime}=N_1^{\prime}-N_5^{\prime}$. 

In order to understand better the F1/NS5 sector
of this black hole we also postulate that there exists a radius given by $r_b$ at which 
a shell composed of  $\delta N_{NS}$ NS5-branes and $\delta N_{F1}$ F1-strings resides.
The black hole inside of this radius (and inside of $r_c$) therefore contains  
$N_{NS}^{\prime}=N_{NS}-\del N_{NS}$
NS5-branes and $N_{F1}^{\prime}=N_{F1}-\del N_{F1}$ fundamental strings which corresponds
to asymptotic charges $Q_{NS}^{\prime}=N_{NS}^{\prime}$ and  
$Q_{F1}^{\prime}=N_{F1}^{\prime}-N_{NS}^{\prime}$. 

It is important to note that if there is no enhan\c{c}on effect for NS5-branes the
shell located at $r=r_b$ is completely benign and there is no reason that the 
F1/NS5-branes located there cannot be placed inside of the black hole. It will be shown
however that enhan\c{c}on physics is relevant and there is a radius inside of which 
NS5-branes may not proceed unless suitable conditions, which will be derived below,
are satisfied.   

As a last point, note that the relative magnitudes of $r_c$ and $r_b$ are, 
for now, unimportant
because of the BPS property and the fact 
that the curvature of K3 does not 
induce any F1-string charge on the D5-branes nor any D1-brane 
charge on the NS5-branes.
In effect the two shells are independent.

\subsection{D1/D5-brane Enhan\c{c}ons}

The more familiar case of the D1/D5 enhan\c{c}on is considered first. Inside
the radius $r_c$ the supergravity solution in eqn.~\reef{10dmet} 
must be modified to account for the different charges. This new solution must however
match onto the exterior solution in a smooth way. This is achieved by simply replacing the
harmonic functions related to the D1-branes and D5-branes in~\reef{10dmet} by the 
following `hatted' functions,
\beq
\hat{h}_1=1+\frac{c_1Q_1^{\prime}}{r}+\frac{c_1(Q_1-Q_1^{\prime})}{r_c}
\,\,\,\,\,\,\,\hat{h}_5=1+\frac{c_5Q_5^{\prime}}{r}+\frac{c_5(Q_5-Q_5^{\prime})}{r_c}.
\labell{rcharm}
\eeq
The NS5-brane and F1-string harmonic functions are unchanged and do not 
figure in the enhan\c{c}on effect for the D1/D5 system. 
Even though this patched metric is continuous at $r=r_c$ (the `hatted' functions were 
chosen to ensure this) there will be a discontinuity in the extrinsic curvature
of the gluing surface $r=r_c$~\cite{junc}. This has an interpretation 
as the stress energy of the D1/D5-brane shell---for a complete discussion see 
ref.~\cite{jmpr}.
The extrinsic curvature of the $r=r_c$ surface is,
\beq
K^{\pm}_{AB}=\frac{1}{2}n_{\pm}^C\partial_{C}g_{AB}|_{r=r_c}
=\mp\frac{1}{2}\sigma\partial_{r}g_{AB}|_{r=r_c}
\labell{excurv}
\eeq
where $n_{\pm}=\mp\sigma\partial_r$ is the outward directed unit normal vector with 
$\sigma=1/\sqrt{g_{rr}}|_{r=r_c}$. 
The discontinuity of the
extrinsic curvature across the surface $r=r_c$ is defined to be 
$\gamma_{AB}=K^{+}_{AB}+K^{-}_{AB}$
and the surface stress tensor is,
\beq
S_{AB}=\frac{1}{8\pi G}\left(\gamma_{AB}-g_{AB}\gamma^{C}{}_{C}\right)
\labell{stress}
\eeq
where $16\pi G=(2\pi)^7g^2l_s^8$ is the ten dimensional Newton's constant.

Calculating the stress tensor of the surface at $r=r_c$ one finds,
\beqa
8\pi GS_{tt}&=&\frac{\sigma}{2}\left(\frac{h_5^{\prime}}{h_5}
+\frac{h_1^{\prime}}{h_1}
-\frac{\hat{h}_5^{\prime}}{\hat{h}_5}-
\frac{\hat{h}_1^{\prime}}{\hat{h}_1}\right)g_{tt}
\nonumber \\
8\pi GS_{44}&=&\frac{\sigma}{2}\left(\frac{h_5^{\prime}}{h_5}
+\frac{h_1^{\prime}}{h_1}
-\frac{\hat{h}_5^{\prime}}{\hat{h}_5}-
\frac{\hat{h}_1^{\prime}}{\hat{h}_1}\right)g_{44}
\nonumber \\
8\pi GS_{99}&=&0
\nonumber \\
8\pi GS_{ab}&=&\frac{\sigma}{2}\left(\frac{h_5^{\prime}}{h_5}-
\frac{\hat{h}_5^{\prime}}{\hat{h}_5}\right)g_{ab}
\nonumber \\
8\pi GS_{ij}&=&0
\labell{rcstress}
\eeqa
where the indices $a,b$ run over the K3 and $i,j$ denote the angular coordinates
on the two-sphere at the incision point $r_c$.
This result is exactly as expected. The 
D5-branes do not wrap the $x^9$ circle so there is no stress-energy in this direction.
Also, since the D1-branes do not wrap the K3 manifold the stress tensor in these 
directions is independent of $h_1$ and $\hat{h}_1$. Finally, since this is a BPS
configuration, there is no stress-energy associated with placing the D5-branes and 
D1-branes on the shell at $r=r_c$.
This result is also manifestly independent of the NS5-branes and fundamental
strings. 

The energy density, or tension, of the effective string in the
$x^4$ direction is given by,
\beq
8\pi GT=\frac{\sigma}{2}\left(\frac{\hat{h}_1^{\prime}}{\hat{h}_1}+
\frac{\hat{h}_5^{\prime}}{\hat{h}_5}
-\frac{h_1^{\prime}}{h_1}-\frac{h_5^{\prime}}{h_5}\right)
\labell{ener2}
\eeq
This expression can be rewritten as,
\beq
T =\frac{h_1^{-1/4}h_5^{1/4}h_{F1}^{1/4}h_{NS}^{-1/4}(Q_1-Q_1^{\prime})\tau_1}
{A_{S^2}A_{K3}A_{S_9^1}}+
\frac{h_1^{1/4}h_5^{-1/4}h_{F1}^{-1/4}h_{NS}^{1/4}(Q_5-Q_5^{\prime})\tau_5}
{A_{S^2}A_{S_9^1}}
\labell{T1T5}
\eeq
where $\tau_{1,5}$ are the canonical flat space tensions of D1--branes and 
D5--branes respectively.
Here $A_{K3},A_{S^2},A_{S_9^1}$ are the proper areas, as measured
at $r=r_c$ in the Einstein frame, of the 
K3 manifold, $S^2$ and the $x^9$ circle respectively. These factors appear in 
the formula for the tension since, from the supergravity point of view, the branes
involved have been smeared over these directions and hence the tension appears as an
average over these directions.
Also, $Q_1-Q_1^{\prime}=
\del N_1 -\del N_5$ and $Q_5-Q_5^{\prime}=\del N_5$ are the effective number of 
D1-branes and D5-branes residing on the shell.
In order to understand this 
result one must recall that the calculations presented here have been carried out 
using the Einstein frame metric. In string frame the {\it local} value of the 
tension of a D1-brane, for example, can be obtained by varying the
the Born-Infeld action,
\beq
S=\tau_{1}\int_{WV} e^{-\phi}\sqrt{P[g_{AB}^{string}]}.
\labell{dbi}
\eeq
with respect to the string frame metric. The result is $T_1=e^{-\phi}\tau_{1}$. 
Switching to Einstein frame using $g^{string}_{AB}=e^{\phi/2}g^{Einstein}_{AB}$ 
and varying with respect to the Einstein frame metric one finds 
that the {\it local} value of the D1-brane tension is,
\beq
T_{1}^{Einstein}=e^{-\phi/2}\tau_{1}=
h_1^{-1/4}h_5^{1/4}h_{F1}^{1/4}h_{NS}^{-1/4}\tau_1
\labell{Tein}
\eeq
which is precisely the result obtained in eqn.~\reef{T1T5} above. Similar 
considerations for the D5-brane verify that eqn.~\reef{T1T5} is the 
expected formula for the effective tension in the string.

Substituting the harmonic functions of eqns.~\reef{4dharm} and~\reef{rcharm}
into eqns.~\reef{ener2} or~\reef{T1T5} it is found that the tension of 
the effective string remains {\it non-negative} so long as,
\beq
r_c \geq \frac{c_5c_1}{\del N_5(c_5-c_1)-\del N_1c_1}\left(\del N_5(2N_5-N_1)-
\del N_1 N_5\right)
\labell{genenh}
\eeq
Further if there are no 
D1-branes on the shell, \ie $\del N_1=0$, then one finds that, 
\beqa
r_c&\geq&\frac{c_5c_1}{c_5-c_1}\left(2N_5-N_1\right)
\nonumber \\
&=&\frac{gl_s^2}{2R_9}\frac{\left(2N_5-N_1\right)}{V/V^{\star}-1}\equiv r_e
\labell{rc}
\eeqa
which is precisely the enhan\c{c}on radius originally found in ref.~\cite{jpp}
rewritten in terms of the four dimensional parameters given in eqn.~\reef{4const}.

From eqn.~\reef{rc} it is obvious that for $2N_5 > N_1$ the tension 
of the D5-brane shell vanishes outside of
the horizon, \ie $r_e > 0$. Inside of this radius the tension would 
be negative and as emphasized
in ref.~\cite{jpp} supersymmetry would be broken. Therefore all of the D5-branes on
the shell must remain outside of the black hole. In particular they cannot contribute to
the entropy.
Note, however, that for $\del N_1 > 0$ 
we can have  $r_c < r_e$ and still maintain non-negative tension in the shell. 
Hence when explicit D1-branes are included on the shell it can be brought closer to the
horizon. Each additional D1-brane on the shell brings the minimum radius infinitesimally
closer to $r=0$.  The enhan\c{c}on thus
belies an onion like structure. This will be seen in more detail in the following section
where it will be shown that probes which are made up of both D1- and D5-branes can move
closer to the horizon, while maintaining positive tension, than D5-branes alone. 
As we shall see in some cases such combined probes may even reach $r=0$.  

Finally, it is straight forward to check that 
the string frame volume of the K3 defined
in eqn.~\reef{volK3} satisfies $V(r_e)=V^{\star}$.

\subsection{F1/NS5-brane Enhan\c{c}ons}

Next, consider the shell of F1/NS5-branes sitting at $r=r_b$. Inside of this radius 
the metric is matched onto the metric of eqn.~\reef{10dmet} where the harmonic 
functions of eqn.~\reef{4dharm} are used for the D1/D5-branes\footnote{In 
the case that $r_b<r_c$ one may wish to use the harmonic functions
given in eqn.~\reef{rcharm} to represent the D1/D5-brane charges in the black hole.
The difference is inconsequential since  these harmonic 
functions do not change form as
one crosses the surface $r=r_b$ and hence whichever set of harmonic functions are 
chosen will drop out of the stress tensor calculation.}
and those for the NS5-branes and F1-strings are given by,
\beq
\hat{h}_{F1}=1+\frac{c_{F1}Q_{F1}^{\prime}}{r}+\frac{c_{F1}(Q_{F1}-
Q_{F1}^{\prime})}{r_b}
\,\,\,\,\,\,\,\hat{h}_{NS}=1+\frac{c_{NS}Q_{NS}^{\prime}}{r}+\frac{c_{NS}(Q_{NS}-
Q_{NS}^{\prime})}{r_b}
\labell{rbharm}
\eeq
Again, the patched metric is continuous but there is a discontinuity in the extrinsic
curvature. Calculating the stress tensor at the surface $r=r_b$ yields,
\beqa
8\pi GS_{tt}&=&\frac{\sigma}{2}\left(\frac{h_{NS}^{\prime}}{h_{NS}}
+\frac{h_{F1}^{\prime}}{h_{F1}}
-\frac{\hat{h}_{NS}^{\prime}}{\hat{h}_{NS}}-
\frac{\hat{h}_{F1}^{\prime}}{\hat{h}_{F1}}\right)g_{tt}
\nonumber \\
8\pi GS_{44}&=&0
\nonumber \\
8\pi GS_{99}&=&\frac{\sigma}{2}\left(\frac{h_{NS}^{\prime}}{h_{NS}}
+\frac{h_{F1}^{\prime}}{h_{F1}}
-\frac{\hat{h}_{NS}^{\prime}}{\hat{h}_{NS}}
-\frac{\hat{h}_{F1}^{\prime}}{\hat{h}_{F1}}\right)g_{44}
\nonumber \\
8\pi GS_{ab}&=&\frac{\sigma}{2}\left(\frac{h_{NS}^{\prime}}{h_{NS}}
-\frac{\hat{h}_{NS}^{\prime}}{\hat{h}_{NS}}\right)g_{ab}
\nonumber \\
8\pi GS_{ij}&=&0
\labell{rbstress}
\eeqa
which, as before is exactly the expected result. In this case there are no branes
wrapping the $x^4$ circle, there are no F1-strings wrapping the K3 and again 
the BPS property of the shell is indicated by the vanishing of the stress 
tensor in the sphere directions. The tension of the effective string lying along
$x^9$ is,
\beq
T=\frac{\sigma}{2}\left(\frac{\hat{h}_{F1}^{\prime}}{\hat{h}_{F1}}+
\frac{\hat{h}_{NS}^{\prime}}{\hat{h}_{NS}}
-\frac{h_{F1}^{\prime}}{h_{F1}}-\frac{h_{NS}^{\prime}}{h_{NS}}\right)
\labell{ener3}
\eeq
as in the previous example this can be rewritten as,
\beq
T =\frac{h_1^{1/4}h_5^{-1/4}h_{F1}^{-1/4}h_{NS}^{1/4}(Q_{F1}-Q_{F1}^{\prime})\tau_{F1}}
{A_{S_4^1}A_{S^2}A_{K3}}+
\frac{h_1^{-1/4}h_5^{1/4}h_{F1}^{1/4}h_{NS}^{-1/4}(Q_{NS}-Q_{NS}^{\prime})\tau_{NS}}
{A_{S^2}A_{S_4^1}}
\labell{TF1TNS}
\eeq
where $\tau_{F1,NS}$ are the canonical flat space tensions of F1-strings and 
NS5-branes respectively.
Here $A_{K3},A_{S^2}$, are the proper areas introduced above and $A_{S_4^1}$ is the
proper area of the $x^4$ circle.
All of these are now  measured
at $r=r_b$ in the Einstein frame. Just as before these factors appear in 
the formula for the tension because the branes
involved have been smeared over these directions and our calculations have
been performed using the Einstein frame metric. In particular varying the
world volume action for F1-strings or NS5-branes with respect to the
Einstein frame metric yields precisely the local value of the tensions found in 
eqn.~\reef{TF1TNS} ---- see section 2.1 above.
Also, $Q_{F1}-Q_{F1}^{\prime}=
\del N_{F1} -\del N_{NS}$ and $Q_{NS}-Q_{NS}^{\prime}=\del N_{NS}$ 
are the effective numbers of 
F1-strings and NS5-branes residing on the shell.
Positivity of this tension requires,
\beq
r_b\geq\frac{c_{NS}c_{F1}}{\del N_{NS}(c_{NS}-c_{F1})-\del N_{F1}c_{F1}}
\left(\del N_{NS}(2N_{NS}-N_{F1})-\del N_{F1} N_{NS}\right)
\labell{genenh2}
\eeq
which for $\del N_{F1}=0$ can be evaluated as,
\beqa
r_b&\geq &\frac{c_{NS}c_{F1}}{c_{NS}-c_{F1}}\left(2N_{NS}-N_{F1}\right)
\nonumber \\
&=&\frac{g^2l_s^2}{2R_4}\frac{\left(2N_{NS}-
N_{F1}\right)}{V/V^{\star}-g^2}\equiv \tilde{r}_e
\labell{rb}
\eeqa
Again the tension in the shell vanishes outside of the horizon for
$2N_{NS}>N_{F1}$ and moving individual NS5-branes off the shell 
towards the horizon will lead to negative tensions and unphysical results.
As in the previous section there is an onion structure since adding explicit 
F1-strings to the shell allows it to be moved closer to the horizon while maintaining 
supersymmetry.

Since in refs.~\cite{jpp,jmpr,5d} tensionless shells were found to coincide
precisely with tensionless supersymmetric probes of the geometry giving rise to 
the enhan\c{c}on effect one should interpret the results here as indicating
the presence of an enhan\c{c}on effect for NS5-branes. 
Further evidence for NS5-brane enhan\c{c}ons will be presented in 
the following section.

It is interesting to note that the form of eqn.~\reef{genenh} 
is identical to that of eqn.~\reef{genenh2} 
up to the replacement of D1/D5 labels with those of F1/NS5. This is exactly the
prescription given by S-duality\footnote{Recall that under S-duality
 $g\rightarrow 1/g$ and 
$l_s^2\rightarrow gl_s^2$.}  and thus these two radii, as well as the two 
effective strings, are interchanged under type IIB S-duality. More will be said about
this, as well as the implications of T-duality, in the following sections. 

As a final comment, observe that in eqn.~\reef{rb} 
there is an additional factor of the asymptotic string coupling $g$ relative to
 eqn.~\reef{rc}. Thus (for weak coupling) the enhan\c{c}on shell for NS5-branes 
sits at a radius which is parametrically smaller than that for the D5-brane.

\section{Probes and Enhan\c{c}ons}

In this section the black hole discussed above from the point of view of pure
supergravity will be analyzed from a more stringy perspective
by probing the black hole geometry using D1/D5/F1/NS5-branes. These are all
natural probes to use since, when aligned correctly, they do not break any
additional supersymmetry. Of course the probe actions for D1-branes, D5-branes and
fundamental strings are all well known, however, the action for the NS5-brane is 
more complicated and it will be obtained here by considering S-duality of the
D5-brane action.

\subsection{D5-brane Probes and S-Duality}

To begin, consider the probe Born-Infeld action for an 
unwrapped D5-brane~\cite{bin}, 
\beq
S_{D5}=-\tau_{D5}\int d^6ye^{-\phi}\sqrt{-det\,P[g_{AB}^{string}]}
\labell{probe1}
\eeq
where $\tau_{D5}=(2\pi)^{-5}g^{-1}l_s^{-6}$ and $P[g_{AB}^{string}]$ is the 
pull-back of the 
background string frame metric to the six dimensional world volume. 
In order to act with S-duality on this action
it is convenient to work with the Einstein frame metric, which is invariant
under S-duality and defined by, $g^{Einstein}_{AB}=e^{-\phi/2}g^{string}_{AB}$. 
The only
field which transforms under S-duality is then the dilaton, $\phi\rightarrow
-\phi$. The result after transforming back to string frame is,
\beq
S_{NS}=-\tau_{NS}\int d^6y\, e^{-2\phi}\sqrt{-det\,P[g_{AB}^{string}]}
\labell{probe2}
\eeq
where $\tau_{NS}=\tau_{D5}/g$. 
In order to determine the probe action for a wrapped NS5-brane we recall the
analogous action for a D5-brane wrapped on K3~\cite{curvcoup,ghm,bbg,jpp},
\beq
S_{D5}=-\int d^2\xi\,e^{-\phi}\left(\tau_{5}V(r)-\tau_{D1}\right)
\sqrt{-det\,P[g_{AB}^{string}]}
+\tau_{D5}\int_{K3\times M}C_{(6)}-\tau_{D1}\int_{M}C_{(2)}
\labell{probe2a}
\eeq

Considering that S-duality interchanges D1-branes and fundamental strings and
recalling that the probe action for a fundamental string is 
simply given by the 
Nambu-Goto action we can now write down the probe action for an NS5-brane
wrapped on a K3 surface and use this to probe the metric of the
F1/NS5 bound state given in eqn.~\reef{NSF1}. The probe action is,
\beq
S_{NS}=-\int d^2\zeta\left(\tau_{NS}e^{-2\phi}V(r)-\tau_{F1}\right)
\sqrt{-det\,P[g_{AB}^{string}]}
+\tau_{NS}\int_{K3\times M}B_{(6)}-\tau_{F1}\int_{M}B_{(2)}
\labell{probe3}
\eeq
This has been written down on physical grounds dictated by S-duality and 
supersymmetry. In this expression $\tau_{F1}=(2\pi l_s^2)^{-1}$, $M$ represents
the unwrapped directions of the NS5-brane's world volume and 
$V(r)$ is the local volume of the K3 surface. Note that standard supergravity
conventions for the tensions and charges, as discussed in ref.~\cite{dielec},
have been employed in writing down these actions.

\subsection{Probing the Black Hole}

In this section the probe technology above will be applied to the four dimensional
black hole discussed previously. This analysis is most easily performed using the 
string frame metric for this black hole which is given by,
\beqa
ds_{string}^2&=&-h_1^{-1/2}h_5^{-1/2}h_{F1}^{-1}dt^2 + h_1^{-1/2}h_5^{-1/2}h_{NS}dx_4^2
+ h_1^{1/2}h_5^{1/2}h_{F1}^{-1}dx_9^2 
\nonumber \\
&+& h_1^{1/2}h_5^{-1/2}ds_{K3}^2 
+h_1^{1/2}h_5^{1/2}h_{NS}\left(dr^2+r^2d\Omega_2^2\right)
\labell{10dmet2}
\eeqa
while the remainder of the closed string fields are the same as in eqn.~\reef{bhdil}. 
We will probe the metric given in eqn.~\reef{10dmet2} 
using all of the constituents of the black hole. 
In order to make a direct comparison with the supergravity calculations above
the probes used in this section will be made up of, either D1/D5 bound states
involving $n_5$ D5-branes and $n_1$ D1-branes or F1/NS5 bound states composed of
$n_{NS}$ NS5-branes and $n_{F1}$ fundamental strings. It should be emphasized that
these will be considered as true bound states in the sense that the D1-branes exist
inside the world volume of the D5-branes as finite size instantons. 
This implies that the strings which describe the separation
of the D1-branes from the D5-branes in the Coulomb phase are all massive. Similar 
comments apply for the F1-strings embedded into the NS5-brane world volume.

The effective Lagrangians are found by calculating the pull back of 
the ten dimensional 
metric in ~\reef{10dmet2} to the world volumes of the probes and expanding in
powers of $v^2$~\cite{dps}. Note that the world volumes are {\it not} the same. The 
D1/D5 probe, in static gauge, is aligned along the 
$t,x^4$ directions while the
F1/NS5 probe is aligned in the $t,x^9$ plane.

We will use the D5-brane probe action in eqn.~\reef{probe2a} for $n_5$ D5-branes
and include $n_1$ D1-branes as instantons on the world volume 
so that $q_1\tau_1=(n_1-n_5)\tau_1$ is the effective total
D1-brane charge. We will also be working in static 
gauge where,
\beq
\xi^0=t, \,\,\,\,\,\xi^1=x^4, \,\,\,\,\, r=r(t,x^4), \,\,\,\,\, \theta=\theta(t,x^4), 
\,\,\,\,\, \phi=\phi(t,x^4)
\labell{static1}
\eeq
and the probe is frozen on the K3 as well as the $x^9$ circle.
The kinetic term in the probe action~\reef{probe2a} is found to be,
\beq
{\cal L}=\frac{1}{2}\left(n_5\tau_5Vh_1+(n_1-n_5)\tau_1h_5\right)v^2
\labell{probeten1}
\eeq
where the the `velocity' is given by,
\beq
v^2=h_{F1}h_{NS}\left[\dot{r}^2 + r^2\dot{\Omega}_2^2-
h_{F1}^{-1}h_{NS}^{-1}\left({r^{\prime}}^2+r^2{\Omega_2^{\prime}}^2\right)\right]
\labell{D1D5vel}
\eeq
here `dot' and `prime' denote differentiation with respect to $t$ and $x^4$ 
respectively. Also we have used the shorthand notation 
$\dot{\Omega}_2^2\equiv \dot{\theta}^2+\sin^2\theta\,\dot{\phi}^2$ and likewise for
${\Omega_2^{\prime}}^2$.
The effective probe tension is given by the pre-factor in eqn.~\reef{probeten1}
and due to the relative sign 
appearing there it is expected to become negative at some radius for $n_5 > n_1$. 
Requiring that the tension always remain positive imposes that,
\beq  
r \geq \frac{c_5c_1}{n_5(c_5-c_1)-n_1c_1}\left(n_5(2N_5-N_1)-
n_1 N_5\right)
\labell{minrad1}
\eeq
which, upon replacing $n_5\rightarrow \del N_5$ and  $n_1\rightarrow \del N_1$, 
is the {\it same} requirement found from the supergravity calculations.
As pointed out in ref.~\cite{jmpr,5d} this indicates that there is complete consistency
between the probe and supergravity approaches.
There are three special cases of this result. First, notice that setting $n_5=0$ in
these expressions, so that the probe is composed entirely of D1-branes, gives,
\beq
{\cal L}=\frac{1}{2}n_1\tau_1h_5v^2
\labell{n5is0}
\eeq
and the effective tension is everywhere positive. This is simply the obvious
result that D1-branes are not effected by compactification on a 
transverse K3 manifold and hence do not see any enhan\c{c}on~\cite{cvjtalk}.
Next, setting $n_1=n_5$, so that the D5-branes are `dressed' with the appropriate
number of instantons to effectively cancel the D1-brane charge induced by the 
wrapping on K3, eqn.~\reef{probeten1} reduces to,
\beq
{\cal L}=\frac{1}{2}n_5\tau_5h_1v^2
\labell{n5isn1}
\eeq 
which is also everywhere positive. This implies that a D5-brane probe can
in fact move past the naively minimum radius in eqn.~\reef{minrad1} and into the
black hole so long as it
is accompanied by the appropriate number of bound D1-branes.
Finally, if there are no D1-branes bound to the D5-brane, \ie $n_1=0$ then
the D5-branes cannot access the region of the space time inside of,
\beqa
r&=&\frac{c_5c_1}{c_5-c_1}\left(2N_5-N_1\right)
\\ \nonumber
&=&\frac{gl_s^2}{2R_9}\frac{\left(2N_5-N_1\right)}{V/V^{\star}-1}=r_e
\labell{probeenha}
\eeqa
which agrees exactly with the supergravity result in eqn.~\reef{rc}.
As remarked above this is simply the enhan\c{c}on radius originally found in 
ref.~\cite{jpp} rewritten in terms of parameters appropriate to four dimensions.

The above results generalize nicely to the NS5-brane probe. We choose static gauge
so that the probe is aligned along
the $x^9$ direction, \ie  
\beq
\zeta^0=t, \,\,\,\,\, \zeta^1=x^9, \,\,\,\,\, r=r(t,x^9), \,\,\,\,\, 
\theta=\theta(t,x^9), \,\,\,\,\, \phi=\phi(t,x^9)
\labell{static2}
\eeq
as well as being frozen on the K3 and $x^4$ circle. The probe 
action found in eqn.~\reef{probe3} for $n_{NS}$ NS5-branes with 
$n_{F1}$ bound F1-strings yields the following kinetic term,
\beq
{\cal L}=\frac{1}{2}\left(n_{NS}\tau_{NS}Vh_{F1}+(n_{F1}-n_{NS})\tau_{F1}h_{NS}
\right)v^2
\labell{probeten2}
\eeq
where the the `velocity' is now given by,
\beq
v^2=h_{1}h_{5}\left[\dot{r}^2 + r^2\dot{\Omega}_2^2-
h_{1}^{-1}h_{5}^{-1}\left({r^{\prime}}^2+r^2{\Omega_2^{\prime}}^2\right)\right].
\labell{F1NSvel}
\eeq
The probe tension remains positive as long as,
\beq  
r \geq \frac{c_{NS}c_{F1}}{n_{NS}(c_{NS}-c_{F1})-n_{F1}c_{F1}}\left(n_{NS}(2N_{NS}-
N_{F1})-n_{F1} N_{NS}\right)
\labell{minrad2}
\eeq
These results are, of course, the same as those for the D1/D5 probe above up to the
exchange of D1/D5 labels for F1/NS5 labels. 
Again there are three interesting special cases. When $n_{NS}=0$ 
eqn.~\reef{probeten2} collapses to the simple case of $n_{F1}$ fundamental strings
probing the black hole,
\beq
{\cal L}=\frac{1}{2}n_{F1}\tau_{F1}h_{NS}v^2.
\labell{nNSis0}
\eeq
The tension is everywhere positive and, just as for the
solitary D1-branes above, these fundamental strings do not see any enhan\c{c}on 
shell. When $n_{F1}=n_{NS}$ so that the fundamental string charge, induced by 
wrapping the NS5-branes on the K3, is completely canceled,
\beq
{\cal L}=\frac{1}{2}n_{NS}\tau_{NS}h_{F1}v^2
\labell{nNSisnF1}
\eeq 
and the probe tension is again
always positive. This `dressed' NS5-brane does not see any enhan\c{c}on
and can thus be positioned at any arbitrary radius, in particular it can
be moved inside the black hole.
Finally, if $n_{F1}=0$ then the NS5-brane probe cannot maintain positive tension 
unless it remains outside the enhan\c{c}on radius,
\beqa
r&=&\frac{c_{NS}c_{F1}}{c_{NS}-c_{F1}}\left(2N_{NS}-N_{F1}\right)
\\ \nonumber
&=&\frac{g^2l_s^2}{2R_4}\frac{\left(2N_{NS}-
N_{F1}\right)}{V/V^{\star}-g^2}=\tilde{r}_e
\labell{probeenhb}
\eeqa  

Again this in precise agreement with the supergravity calculations of the previous 
section.

In the next section we explore the implications of the results presented above
for black hole thermodynamics.

\section{Entropy and the Enhan\c{c}on}

\subsection{Aspects of the Black Hole}

To understand how the results obtained thus far are relevant for 
black hole physics it is 
instructive to attempt to build our black hole from its constituents.
Consider a system of $N_1,N_5,N_{F1},N_{NS}$ D1-branes, D5-branes, F1-strings
and NS5-branes respectively. To begin these are all assumed to be sitting 
at asymptotic infinity.
Next, consider placing $N_1^{\prime}$ D1-branes and $N_{F1}^{\prime}$ F1-strings
at the origin. In order to ensure that the local string coupling remains small
in what follows we assume that $N_1^{\prime} \ll N_{F1}^{\prime}$. 
The probe analysis of the previous section indicates that there is no obstacle
to moving $N_5^{\prime}$ D5-branes and $N_{NS}^{\prime}$ NS5-branes to 
the origin to form a black hole so long as $2N_5^{\prime}<N_1^{\prime}$ and
$2N_{NS}^{\prime}<N_{F1}^{\prime}$. The rest of the wrapped branes can come no 
closer than their respective enhan\c{c}on shells located at $r_e$ and 
$\tilde{r}_e$ given in eqns.~\reef{rc},\reef{rb}. From this perspective 
it would seem that the limiting configuration, in which both of the enhan\c{con}
shells sit exactly at the horizon, involves all of the available
D1-branes and F1-strings while only utilizing $N_5^{\prime}=N_1/2$ 
D5-branes and $N_{NS}^{\prime}=N_{F1}/2$ NS5-branes. 
The results of the
probe calculations above---see eqns.~\reef{n5isn1},\reef{nNSisnF1}---
indicate however that when a D5-brane (NS5-brane) is suitably `dressed'
with a D1-brane (F1-string) that its tension will always be positive. Such a
bound state is therefore free to proceed into the black hole. In this way 
one may construct a 
black hole with any values for the asymptotic charges by continually adding
dressed D5- or NS5-branes until one either runs out of D5- or NS5-branes or all 
of the available D1-branes and F1-strings have been used up. 

For each set of branes (D1/D5 or F1/NS5) there are three possibilities. As an
illustration consider only the D1/D5-brane sector. The exact same comments will
apply to the F1/NS5 sector.
First, when $0<N_1<N_5$ the black hole can only contain up to $N_5^{\prime}=N_1$
D5-branes. The remaining D5-branes must remain at the enhan\c{c}on radius since 
there are no more D1-branes with which to dress them. Second, if $N_5<N_1<2N_5$
then the black hole can absorb all of the D5-branes. In this case however 
there is still a
region outside of the horizon where an additional D5-brane probe would 
have negative tension since
from eqn.~\reef{rc} the enhan\c{c}on radius is still positive. 
Finally if $2N_5<N_1$ then all of the D5-branes can clearly be absorbed
and there is no enhan\c{c}on appearing outside of the horizon so that a D5-brane 
probe of the geometry set up by this final configuration 
could be brought from infinity and fall through the horizon without encountering
any difficulty associated with enhan\c{c}ons.
Of course, all of the comments made here apply directly to the F1/NS5 components of
this black hole as well. 

The supergravity solution which describes the second and third scenarios is simply
given by eqn.~\reef{10dmet} taken in the range $0<r<\infty$. This is because
all of the available branes can be used in the construction of the black hole. In the
first case the excess D5- or NS5-branes must reside at their respective 
enhan\c{c}on radii outside the horizon. The solution therefore is given by 
a patched metric as discussed in section 2 where the `hatted' harmonic functions,
appearing in eqns.~\reef{rcharm},\reef{rbharm}, are used to describe the 
geometry interior to the enhan\c{c}on loci while the `unhatted'
harmonic functions describe the geometry exterior to the enhan\c{c}on loci.  
It should be clear that it is possible for both of the D5-brane and NS5-brane 
enhan\c{c}on radii to be positive, or for both to be negative or a combination of the
two. In other words all combinations of the hatted and unhatted harmonic 
functions are possible.

The conclusion to be drawn from this discussion is that enhan\c{c}on physics
is relevant for these black holes when the number of D1-branes (F1-strings) 
is small compared to the number of D5-branes (NS5-branes). In particular the 
enhan\c{c}on mechanism places an upper limit on the number of D5-branes 
(NS5-branes) that can be used to build the black hole when there are a finite number
of D1-branes (F1-strings) available.
As was shown for five dimensional black holes in ref.~\cite{5d} and as will 
be shown in the next section, this facet of enhan\c{c}on physics 
is absolutely crucial for ensuring that this class of 
black holes are not in direct conflict with the second law of thermodynamics. 

Finally, it should be emphasized that any excess D5-branes or NS5-branes that cannot 
be brought into the black hole, in the case $0<N_1<N_5$ for example, can either
be left on the enhan\c{c}on shell or moved off to infinity and removed from the
problem. The area of the horizon, and thus the entropy of the black hole, 
is identical in either case.

\subsection{The Second Law and the Enhan\c{c}on}
 
The Bekenstein-Hawking entropy for the black holes considered here is given by,
\beq
S=\frac{A}{4G}=2\pi\sqrt{(N_1-N_5)(N_{F1}-N_{NS})N_5N_{NS}}.
\labell{bhent}
\eeq
As in the case of five dimensional black holes\cite{5d} it is instructive to consider
how this quantity depends on the numbers of D5-branes and NS5-branes for a fixed 
quantity of D1-branes and F1-strings. As can be seen in fig.~\reef{fig:area} this
is described by a half-ellipsoid. It is clear from this plot that in order to
maximize the entropy the black hole should be constructed from precisely $N_5=N_1/2$
D5-branes and $N_{NS}=N_{F1}/2$ NS5-branes. Intriguingly when there are only a small
number of available D1-branes and/or fundamental strings the entropy is maximized
by utilizing only a fraction of the available D5- or NS5-branes. We will comment on 
this further in the discussion section. For now however note that this maximum entropy 
black hole occurs exactly when {\it both} enhan\c{c}on radii coincide with 
the horizon at $r=0$.
\setlength{\unitlength}{1mm}
\begin{figure}[ht!]
\centerline{\includegraphics[width=8.5cm]{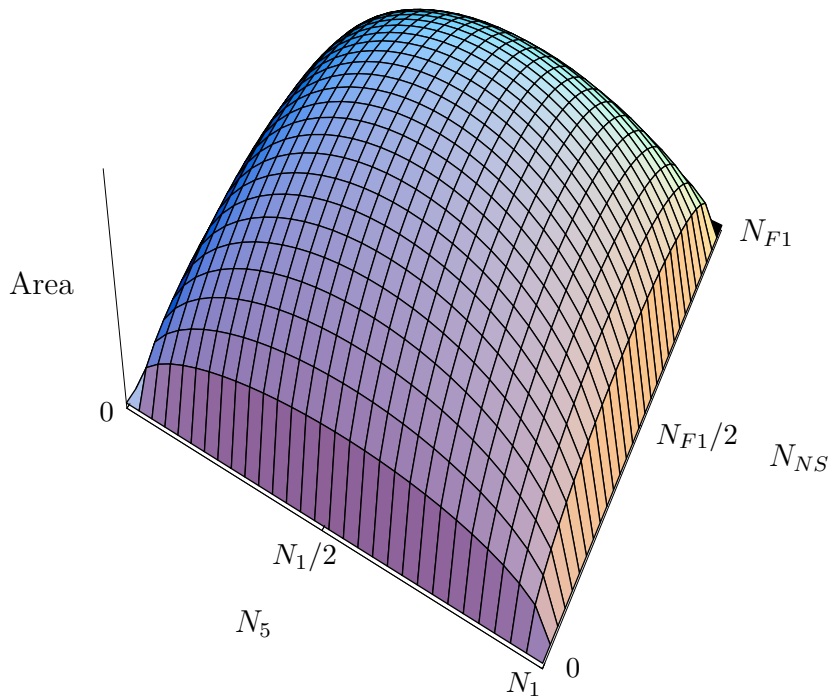}}
\begin{picture}(0,0)
\put(30,55){Area}
\put(65,19){\small $N_1/2$}
\put(96,2){\small $N_1$}
\put(60,10){$N_5$}
\put(116,35){\small $N_{F1}/2$}
\put(127,62){$N_{F1}$}
\put(131,32){$N_{NS}$}
\put(42,38){\small 0}
\put(104,4){\small 0}
\end{picture}
\caption{\small The area of the black hole horizon as a function of $N_5$ and $N_{NS}$
for fixed $N_1$ and $N_{F1}$. Maximizing the area requires that $N_5=N_1/2$ and
 $N_{NS}=N_{F1}/2$.}
\labell{fig:area}
\end{figure}

To understand the connection between the second law and enhan\c{c}on physics let us
now revisit the construction of this black hole in the previous section. Recall that
we began with a large, but fixed number of D1-branes and F1-strings 
inside the black hole. This places us somewhere near the origin (left corner) of 
fig.~\reef{fig:area}. Now since both $N_1 > 2N_5$ and $N_{F1} > 2N_{NS}$ are 
presumed to be satisfied we can add, freely, either D5-branes or NS5-branes 
and in the process climb towards the apside of the ellipsoid in fig.~\reef{fig:area}.
Now suppose, for example, that we reach  a point at which $N_{NS}=N_{F1}/2$ while 
$N_1 > 2N_5$ continues to be satisfied. We may either add another D5-brane, thereby
continuing our progression towards the absolute maximum, or we may add another
NS5-brane and {\it decrease} the entropy! It should be re-emphasized that 
this system is supersymmetric and additional D5- and NS5-branes may be brought in from
infinity as slowly as one likes \ie adiabatically. Thus adding one more
NS5-brane is clearly a violation of the second law of thermodynamics: Entropy can
never decrease via such a process.

The resolution of this conundrum~\cite{5d} is that precisely when $N_{NS}=N_{F1}/2$
the enhan\c{c}on radius for the NS5-branes is coincident with the 
event horizon and thus our attempt to add an additional NS5-brane into 
the fray is circumvented. 

It is of course possible to bind  NS5-branes with additional fundamental strings
and then allow the bound state to cross the horizon. The 
enhan\c{c}on also prevents probes of this type from lowering the
entropy. Consider attempting to add a probe consisting of $n_{F1}$ fundamental
strings bound to $n_{NS}$ NS5-branes. Assuming these to be small numbers
compared with $N_{F1}$ and $N_{NS}$  it is easy to see from 
eqn.~\reef{bhent} that the resulting change in the (square of the) entropy will be,
\beq
\delta S^2 =4\pi^2(N_1-N_5)N_5\left(n_{NS}(N_{F1}-2N_{NS})+n_{F1}N_{NS}\right)
\labell{varentropy}
\eeq
which is negative precisely when the enhan\c{c}on appears outside of the horizon in
eqn.~\reef{minrad2}! 
The arguments of this section have focussed primarily on the F1/NS5 sector of these
black holes. However, it should be clear that replacing NS5-branes by D5-branes 
and F1-strings by D1-branes in the preceding arguments shows that the enhan\c{c}on
prevents any attempts to violate the second law in this sector of the black hole as
well.

To conclude, a naive interpretation of eqn.~\reef{bhent} indicates that 
by indiscriminately adding D5-branes or NS5-branes to this class of black holes
one could easily violate the second law. As we have seen,
any such attempt is strongly opposed by the enhan\c{c}on.

\section{Duals of the Enhan\c{c}on in five dimensions}

Motivated by the appearance (and necessity) of enhan\c{c}on physics 
for NS5-branes in the context of four dimensional black holes  
we now study the emergence of the enhan\c{c}on
for NS5-branes in isolation. We do so for both the NS5-branes of type IIB and 
type IIA string theory. 

In the original work of~\cite{jpp} the enhan\c{c}on was 
unearthed by considering D6-branes wrapped on K3, or equivalently the T-dual
set up of D5-brane wrapped on K3. Performing a type IIB S-duality transformation
on the latter configuration yields a type IIB NS5-brane (NS5$_{\rm{B}}$) wrapped on
K3. As argued elsewhere in this paper the negative D1-brane charge 
induced by wrapping the D5-branes on K3 transforms under S-duality to give
negative fundamental string charge bound to the world volume of the NS5$_{\rm{B}}$-branes.
As in the specific example of the black hole above this leads to an enhan\c{c}on
locus. This will be briefly demonstrated below via both 
supergravity and probe calculations. 

One may also consider compactifying and performing a T-duality along the 
effective string which 
remains after the NS5$_{\rm{B}}$-branes have been wrapped on K3. This leaves 
NS5$_{\rm{A}}$-branes in type IIA string theory wrapped on a K3 manifold. The
induced fundamental strings which were wound around the (compact) direction on the IIB
side of this T-duality become induced momentum modes propagating on the T-dual circle.
It is tempting to conjecture that this too will yield enhan\c{c}on physics. 
Evidence, via a supergravity calculation, will be presented that this is indeed
the case. 

\subsection{Type IIB NS5-branes and the Enhan\c{c}on}
We begin by considering  the supergravity solution 
for the D1/D5 system of type IIB compactified on K3. 
The Einstein frame solution is,
\beqa
ds^2&=&h_1^{-3/4}h_5^{-1/4}\left(-dt^2+dx_9^2\right)+h_1^{1/4}h_5^{3/4}
\left(dr^2+r^2d\Omega_3^2\right)+h_1^{1/4}h_5^{-1/4}ds_{K3}^2
\nonumber \\
e^{2\phi}&=&\frac{h_1}{h_5}
\nonumber \\
C_{(2)}&=&h_1^{-1}dt\wedge dx^9 
\nonumber \\
C_{(6)}&=&h_5^{-1}dt\wedge dx^5\cdots\wedge dx^9
\labell{D1D5met}
\eeqa
where $ds_{K3}^2$ is the metric on a K3 surface with volume $V$ on which the
directions $x^5,x^6,x^7,x^8$ are wrapped. For later purposes $x^9$ is taken to
be periodic with period $2\pi R_9$.
Also, $h_1$ and $h_5$ are harmonic functions appropriate to five transverse 
directions. 
Performing an S-duality transformation
leaves us in type IIB however the solution~\reef{D1D5met} now becomes that of 
an NS5-brane bound to a fundamental string.
The Einstein frame solution for such an F1/NS5 bound state in type IIB 
wrapped on a K3 surface is given by,
\beq
ds^2=h_{F1}^{-3/4}h_{NS}^{-1/4}\left(-dt^2+dx_9^2\right)+h_{F1}^{1/4}h_{NS}^{3/4}
\left(dr^2+r^2d\Omega_3^2 \right)+ h_{F1}^{1/4}h_{NS}^{-1/4}ds_{K3}^2
\labell{NSF1}
\eeq 
The dilaton and Kalb-Ramond fields are given by,
\beqa
e^{2\phi}&=&\frac{h_{NS}}{h_{F1}}
\nonumber \\
B_{(2)}&=&h_{F1}^{-1}dt\wedge dx^9 
\nonumber \\
B_{(6)}&=&h_{NS}^{-1}dt\wedge dx^5\cdots\wedge dx^9
\labell{dil1}
\eeqa
In these expressions the harmonic functions are,
\beq
h_{F1}=1+\frac{c_{F1}Q_{F1}}{r^2} \,\,\,\,\,\,\, h_{NS}=1+\frac{c_{NS}Q_{NS}}{r^2}.
\labell{harm1}
\eeq  

As in the previous sections we place $\del N_{F1}$ fundamental strings
and $\del N_{NS}$ NS5-branes on a shell located at some radius $R_c$
so that the interior geometry is given by the metric in eqn.~\reef{NSF1}
using the following `hatted' functions,
\beq
\hat{h}_{F1}=1+\frac{c_{F1}Q_{F1}^{\prime}}{r^2}
+\frac{c_{F1}(Q_{F1}-Q_{F1}^{\prime})}{R_c^2}
\,\,\,\,\,\,\, \hat{h}_{NS}=1+\frac{c_{NS}Q_{NS}^{\prime}}{r^2}+
\frac{c_{NS}(Q_{NS}-Q_{NS}^{\prime})}{R_c^2}
\labell{hat}
\eeq
The $c_i$ are now the relevant five dimensional quantities~\cite{juan},
\beq
c_{F1}=g^2l_s^2\frac{V^{\star}}{V} \,\,\,\,\,\,\,c_{NS}=l_s^2
\labell{2bcs}
\eeq
 
Explicit calculation determines the stress tensor at the junction 
surface $r=R_c$ to be,
\beqa
8\pi GS_{\mu\nu}&=&\frac{\sigma}{2}\left(\frac{h_{NS}^{\prime}}{h_{NS}}
+\frac{h_{F1}^{\prime}}{h_{F1}}
-\frac{\hat{h}_{NS}^{\prime}}{\hat{h}_{NS}}-
\frac{\hat{h}_{F1}^{\prime}}{\hat{h}_{F1}}\right)g_{\mu\nu}
\nonumber \\
8\pi GS_{ab}&=&\frac{\sigma}{2}\left(\frac{h_{NS}^{\prime}}{h_{NS}}-
\frac{\hat{h}_{NS}^{\prime}}{\hat{h}_{NS}}\right)g_{ab}
\nonumber \\
8\pi GS_{ij}&=&0
\labell{stress1}
\eeqa
where $\mu,\nu$ are the $t,x_9$ directions, $a,b$ are the K3 directions and
$i,j$ are the three-sphere directions. In this expression $S_{\mu\nu}$ has the
interpretation as the stress-energy in the effective string lying in the 
$t,x^9$ plane. $S_{ab}$ is the stress-energy in the K3 directions which here
is, as expected, coming entirely from the NS5-branes. The stress-energy in the
three-sphere directions vanishes as it should since this is a BPS configuration. 
The tension of the effective string lying in the $x^9$ directions can again
---see section 2---be put in the form expected from varying the Einstein frame world 
volume actions,
\beq
T =\frac{h_{F1}^{-1/4}h_{NS}^{1/4}(Q_{F1}-Q_{F1}^{\prime})\tau_{F1}}
{A_{S^3}A_{K3}}+
\frac{h_{F1}^{1/4}h_{NS}^{-1/4}(Q_{NS}-Q_{NS}^{\prime})\tau_{NS}}
{A_{S^3}}
\labell{TF1TNS2}
\eeq
where $Q_{F1}-Q_{F1}^{\prime} = \del N_{F1}-\del N_{NS}$ and  $Q_{NS}-Q_{NS}^{\prime} 
= \del N_{NS}$.  Substituting the harmonic functions in eqn.~\reef{harm1} this expression is
non-negative only for,
\beq
R_c^2\geq\frac{c_{NS}c_{F1}}{\del N_{NS}(c_{NS}-c_{F1})-\del N_{F1}c_{F1}}
\left(\del N_{NS}(2N_{NS}-N_{F1})-\del N_{F1} N_{NS}\right)
\labell{ns5only}
\eeq
which for $\del N_{F1}=0$ can be evaluated as,
\beqa
R_c^2&\geq &\frac{c_{NS}c_{F1}}{c_{NS}-c_{F1}}\left(2N_{NS}-N_{F1}\right)
\\ \nonumber
&=&\frac{g^2l_s^2}{V/V^{\star}-g^2}\left(2N_{NS}-
N_{F1}\right)\equiv R_e^2.
\labell{Rc}
\eeqa
Note that up to the standard S-duality transformations, $g\rightarrow 1/g$ and
$l_s^2\rightarrow gl_s^2$, this is exactly the enhan\c{c}on locus found in
ref.~\cite{jpp}. The point is that S-duality, and more generally U-duality
simply rotate the charges and the constants $c_i$ into each other.

One can also see the same result emerging by performing a probe calculation
using the action in eqn.~\reef{probe3}. Once again consider a probe composed of
$n_{F1}$ F1-strings which are bound to $n_5$ NS5-branes.
Choosing `static gauge' we align the coordinates $\zeta^1,\zeta^2$ of the 
effective string with the $t,x^9$ direction and allow it to move in the
four non-compact directions. Note that it is frozen on the K3. Explicitly 
calculating the pull-back of eqn.~\reef{NSF1} 
we find that the mass and charge cancel as required by 
supersymmetry and the effective Lagrangian can be written,
\beq
{\cal L}=\frac{1}{2}\left(n_{NS}\tau_{NS}Vh_{F1}+(n_{F1}-n_{NS})\tau_{F1}h_{NS}
\right)v^2
\labell{probeten4}
\eeq
where the the `velocity' is now given by,
\beq
v^2=\left[\dot{r}^2 + r^2\dot{\Omega}_3^2-
{r^{\prime}}^2-r^2{\Omega_3^{\prime}}^2\right].
\labell{F1NSvel2}
\eeq
and the probe tension is physical only for radii satisfying  
eqn.~\reef{ns5only}.

As a final comment we point out that unlike the D1/D5 version of this system the 
enhan\c{c}on radius is not associated with any special behavior of the
K3 surface. In fact the (string frame) volume of 
the K3 is a constant in this metric! 
Note, however that $R_c$ is the point at which the local value of
the string coupling becomes, 
\beq
g^2e^{2\phi}=g^2h_{NS}/h_{F1}=V/V^{\star}
\labell{constant}
\eeq
which is generically a large number. One might therefore worry that string loops
will strongly alter the picture presented here and perhaps negate the
appearance of the enhan\c{c}on within this system. 
There are at least two reasons to believe 
this will not be the case. First, as was demonstrated in section 3, U-duality of
type II string theory allows for the construction of a black hole which, without
enhan\c{c}on physics for NS5-brane's, could easily be used to violate the 
second law of thermodynamics. Further, recall that in this construction the
local dilaton could always be kept small and hence string loops could be neglected.
Second, as was pointed out in ref.~\cite{jpp},
within the enhan\c{c}on locus is a region of enhanced gauge symmetry. It seems
unlikely that this extra gauge symmetry would somehow be destroyed at strong 
coupling. Therefore it ought to be present in any S-dual formulation. The fact that
the enhan\c{c}on radii for the D1/D5 system and the F1/NS5$_{\rm{B}}$ systems
are simply S-dual to each other seems to support this point of view.   

\subsection{Type IIA NS5-branes and the Enhan\c{c}on}

In this subsection we analyze the emergence of the enhan\c{c}on for the case of
NS5$_{\rm{A}}$ branes in type IIA.
Consider performing a T-duality transformation\footnote{Under T-duality 
$R\rightarrow l_s^2/R$ and $g\rightarrow gl_s/R$ when T-duality is performed
on a circle of radius $R$.} on the $x^9$ direction in the metric
~\reef{NSF1}. 
This yields an NS5$_{\rm{A}}$-brane bound 
to momentum propagating in the $x^9$ direction in type IIA string theory. 
It should be emphasized that this momentum is induced by the curvature of K3.
The Einstein frame metric is,
\beq
ds^2=h_5^{-1/4}\left(-dt^2 + dx_5^2 +k\left(dt-dx_5\right)^2\right)+
h_5^{3/4}\left(dr^2+r^2d\Omega_3^2\right)+h_5^{-1/4}ds_{K3}^2
\labell{NSP}
\eeq
while the dilaton and Kalb-Ramond fields are,
\beq
e^{2\phi}=h_{NS} \,\,\,\,\,\,\,B_{(6)}=h_{NS}^{-1}dt\wedge\cdots\wedge dx^9
\labell{dil2}
\eeq
in this expression $h_{NS}$ is identical to that in eqn.~\reef{harm1} while
the `momentum' function is,
\beq
k=\frac{c_pQ_p}{r^2}\,\,\,\,\,\,\,\,\,c_p=\frac{g^2l_s^4}{R_9^2}V^{\star}/V
\labell{harm3}
\eeq
Recall that $Q_p=N_p-N_{NS}$ and $N_{NS}=Q_{NS}$ \ie we have explicitly included
$N_p$ momentum modes propagating along the effective string in addition
to those induced by the curvature.
As in the previous cases we place $\del N_p$ momentum modes and $\del N_{NS}$ 
NS5-branes on a shell located at $r=R_b$. Inside of the shell the harmonic function
for the momentum becomes,
\beq
\hat{k}=\frac{c_pQ_p^{\prime}}{r^2}+\frac{c_p(Q_p-Q_p^{\prime})}{R_b^2}.
\labell{hatted}
\eeq
The harmonic function for the NS5-branes was already given in eqn.~\reef{hat}. 
Calculating the stress tensor of the junction surface yields a more 
complicated expression than before which can be put in the form,
\beqa
8\pi GS_{\mu\nu}&=&\frac{\sigma}{2}\left(\frac{h_{NS}^{\prime}}{h_{NS}}
-\frac{\hat{h}_{NS}^{\prime}}{\hat{h}_{NS}}\right)g_{\mu\nu}+
\frac{\sigma}{2}\left(\hat{k}^{\prime}-k^{\prime}\right)p_{\mu}p_{\nu}
\nonumber \\
8\pi GS_{ab}&=&\frac{\sigma}{2}\left(\frac{h_{NS}^{\prime}}{h_{NS}}
-\frac{\hat{h}_{NS}^{\prime}}{\hat{h}_{NS}}\right)g_{ab}
\nonumber \\
8\pi GS_{ij}&=&0
\labell{stress2}
\eeqa
where $\mu,\nu$ span $t,x^9$ while $a,b$ are indices on the K3 and $i,j$ represent the
angles of the three-sphere at $r=R_b$. 
Here we have also defined the vector 
${\bf p}=h_{NS}^{-1/8}\left(-1,1,{\bf 0}\right)$. 
Note that this is a null vector. The $S_{\mu\nu}$ component  is just 
saying that there is a string with
a tension given by the piece proportional to the metric with a momentum
wave traveling at the speed of light along it.
This is an intuitively pleasing answer but it is also problematic. 
Thus far the procedure has been to determine at which radius the tension of an
effective string vanishes. The vanishing occurs because of competing terms in the
total tension of the string.
Usually one does not think 
about balancing the tension of an extended object with momentum 
to get a zero result, 
it would seem however that this is what is required in this case. In fact one 
can recover the expected enhan\c{c}on radius by setting the energy density, $S_{tt}=0$. In order to alleviate any anxiety regarding this procedure recall 
that above it was
alluded to that the constants appearing in the enhan\c{c}on radius 
are just mixed up by U-duality. This is the five dimensional U-duality group. 
We will therefore perform this analysis in five dimensions where all of the
charges are interpreted as point like sources and are therefore on equal
footing. To do so we
reconsider this bound state in a dimensionally reduced set-up. The string frame
metric in a form suitable for dimensional reduction is,
\beq
ds^2=-\frac{dt^2}{1+k}+(1+k)\left(dx_9-\frac{k}{1+k}dt\right)^2
+h_{NS}\left(dr^2+r^2d\Omega_3^2\right) +ds_{K3}^2
\labell{string}
\eeq
We perform the dimensional reduction on this and shift to the five-dimensional
Einstein frame using, $g_{AB}^{Einstein}=e^{-4\phi_{5}/3}g_{AB}^{string}$ 
where the five dimensional dilaton is given by,
\beq
e^{2\phi_5}=(2\pi R_9V)^{-1}(1+k)^{-1/2}h_{NS}.
\labell{dil5d}
\eeq
The dimensionally reduced Einstein frame metric is,
\beq
ds_5^2=-(h_{NS}(1+k))^{-2/3}dt^2+(h_{NS}(1+k))^{1/3}
\left(dr^2+r^2d\Omega_3^2\right)
\labell{5dE}
\eeq
Applying the same analysis as above we suppose that $\del N_{NS}$ NS5-branes 
and $\del N_p$  momentum modes get held up at 
the radius $R_b$ and imagine matching onto the interior solution with the hatted
functions as above. Calculating the stress tensor at the surface $r=R_b$
now gives,
\beqa
8\pi GS_{tt}&=&\frac{\sigma}{2}\left(\frac{h_{NS}^{\prime}}{h_{NS}}-
\frac{\hat{h}_{NS}^{\prime}}{\hat{h}_{NS}}+
\frac{k^{\prime}}{1+k}
-\frac{\hat{k}^{\prime}}{1+\hat{k}}\right)g_{tt}
\nonumber \\
8\pi GS_{ij}&=&0
\labell{energy}
\eeqa
The energy density is now simply the negative of the factor multiplying $g_{tt}$,
\beq
E=\frac{\sigma}{2}\left(\frac{\hat{k}^{\prime}}{1+\hat{k}}-
\frac{k^{\prime}}{1+k}+\frac{\hat{h}_{NS}^{\prime}}{\hat{h}_{NS}}
-\frac{h_{NS}^{\prime}}{h_{NS}}\right).
\labell{ener}
\eeq
Substituting the harmonic functions the energy
density is non-negative only for,
\beq
R_b^2\geq\frac{c_{NS}c_{P}}{\del N_{NS}(c_{NS}-c_{P})-\del N_{P}c_{P}}
\left(\del N_{NS}(2N_{NS}-N_{P})-\del N_{P} N_{NS}\right)
\labell{ns5only2}
\eeq
and assuming that $\del N_p =0$
\beqa
R_b^2&\geq &\frac{c_{NS}c_{p}}{c_{NS}-c_{p}}(2N_{NS}-N_p)
\nonumber \\ 
&=&g^2l_s^2\frac{2N_{NS}-N_p}{\frac{R_9^2}{l_s^2}V/V^{\star}-g^2}\equiv \tilde{R}_e^2
\labell{rad3}
\eeqa
and thus there is an enhan\c{c}on like effect occurring here as well.
One may note that under the standard T-duality transformations this becomes   
precisely the expression found above for the NS5$_{\rm{B}}$-brane in 
eqn.~\reef{Rc}.

\section{Discussion}

At strong coupling, or in large numbers, Dp-branes 
produce a non-trivial back reaction on the geometry and their effective description
lies within the realm of classical supergravity. 
Therefore the appearance of naked singularities 
and violations of the laws of 
laws of black hole mechanics   
in the gravitational description of Dp-branes wrapped
on K3 manifolds would both pose serious problems. On the one hand the occurrence of
naked singularities casts doubt on whether or not supergravity is in fact 
a consistent description of low energy string theory compactified on K3.
On the other hand violations of the second law of thermodynamics by collections of 
BPS objects in string theory would be a major obstacle for string theory itself.    
Happily both of these undesirable scenarios are prevented by the enhan\c{c}on 
mechanism. It seems therefore that the enhan\c{c}on is itself a fundamental 
aspect of string theory compactifications and one wonders if its influence is not
more ubiquitous. 

Black holes have always provided an interesting arena in which
to explore and gain an understanding of fundamental aspects of quantum gravity and 
string theory. In this paper we have seen how the enhan\c{c}on mechanism of
ref.~\cite{jpp} plays a central role in ensuring that four dimensional
black holes embedded into K3 compactifications of type IIB string theory 
are consistent with the second law of thermodynamics. 

Especially crucial to this result is the fact, implied by S-duality, 
that NS5-branes in K3 compactifications
are subject to the enhan\c{c}on mechanism. Further T-duality, and therefore
U-duality, implies that the same physics is relevant for type IIA NS5-branes as well.
In fact since all of the enhan\c{c}on radii uncovered here were dependent only on 
the constants generically labelled $c_i$ where $i=1,5,F1,NS,P$ it seems that 
the U-duality {\it covariant} expression for the enhan\c{c}on radius, 
analogous to eqns.~\reef{rc},\reef{rb},\reef{Rc},\reef{rad3} for any
stringy object which entirely wraps a K3 manifold can be written in $d$-dimensions as,
\beq
r_e^{d-3}=\frac{c_Ac_b}{c_A-c_b}\left(2N_A-N_b\right).
\labell{gen45}
\eeq
Here the upper case indices represent the constant associated with the wrapped 
object and the lower case indices that of the charged object induced by the curvature
of K3. Establishing the validity of this formula in general would be very interesting.
One immediate implication would be that enhan\c{c}on physics is important
for Kaluza-Klein (KK) monopoles\cite{gross}. Another indication that this is the case
is that U-duality guarantees that the charges making up the
black hole considered in this paper can be rotated to produce a black hole constructed
out of D1-branes, D5-branes, KK monopoles and momentum modes. 
The microscopic entropy of this black hole was considered in refs.\cite{rrc,4d}.
If this black hole were to avoid violating the second law it too would be subject to 
enhan\c{c}on physics for {\it both} the D5-branes as well as the KK monopoles.

Enhan\c{c}on physics for objects other than D-branes raises some curious puzzles. Recall
that in the case of a Dp-brane wrapped on K3 that there is one unit of negative 
charge corresponding to a D(p-4)-brane induced onto the world volume. This does not,
however, correspond to anti-D(p-4)-brane charge. The induced charge preserves
the same supersymmetries as a regular D(p-4)-brane. To be specific consider the D1/D5 
system. In the discussions above $N_1$ D1-branes were bound to $N_5$ D5-branes which were
wrapped on K3 inducing $-N_5$ units of D1-brane charge.  The induced  charge 
is aligned with the 
{\it same} orientation as the `real' D1-branes. Under S-duality this becomes the
F1/NS5 bound state of type IIB string theory with $-N_{NS}$ units of curvature induced 
F1-string charge, again aligned parallel to the `real' F1-strings. As discussed
above T-duality takes this to $N_{NS}$ NS5-branes in IIA bound to $N_p$ units of
right moving momentum and $-N_{NS}$ units of induced momentum which are also right moving.
Here the induced momentum {\bf cannot} be interpreted as left moving momentum 
in the same way as the induced D1-branes that we began with are not anti-D1-branes. 
The induced momentum
preserves the same supersymmetries as the right moving momentum and therefore is itself
right moving. To see this more clearly note that from eqns.~\reef{NSP} 
and~\reef{harm3} the $t,x^5$ components of the metric are,
\beq
ds^2 \propto -dt^2+dx_5^2+\frac{c_p(N_p-N_{NS})}{r^2}(dt-dx_5)^2.
\labell{t5}
\eeq
The fact that it is $dt-dx_5$ which appears and not $dt+dx_5$ indicates that this is
all right moving momentum, even though there are $N_p-N_{NS}$ units of it.     
It seems difficult to understand how this might be incorporated into the  
microscopic entropy of a black hole constructed from a D2/D6/P/NS5 bound state 
wrapped on K3. Usually the contribution of the momentum modes to the degeneracy 
of states is accounted for 
in a partition function. It seems that one would partition only the $N_p$ units of 
`real' momentum since the NS5-brane's contribution enters in the $4Q_2Q_6Q_{NS}$ massless
$(2,6)$ strings which carry the momentum modes. 
Of course this gives the wrong answer since
the asymptotic charge which must appear in the entropy formula is $Q_p=N_p-N_{NS}$. 
The same comments clearly apply to the D1/D5/F1/NS5 black hole considered in this paper
with `momentum' replaced by `winding'. It would be interesting to understand
how the microscopic counting works out in detail.  

The fact that it is entropically favoured for black holes to form which are made up of 
precisely $N_5=N_1/2$ D5-branes and $N_{NS}=N_{F1}/2$ NS5-branes is a very interesting 
result that certainly deserves more investigation. U-duality suggests that these 
relations should be true no matter which four charges make up the black hole and it would
be very useful if one could express these as invariants of the four dimensional U-duality 
group.  Alternatively, insight as to why these particular values of the charges 
are favoured
might be gained from studying slightly non-extremal versions of these black holes. 
Non-extremality would lift the degeneracy of the supersymmetric 
vacua and the dynamics would 
pick a preferred ground state. As discussed in ref.~\cite{5d} this
leads to the intriguing
possibility that D5-branes and NS5-branes can be expelled from these black holes.

Another issue which warrants consideration is whether or not the enhan\c{c}on 
enters into the attractor flow descriptions of black holes in compactified 
supergravity~\cite{kallosh}. It seems probable that recent work of Denef and 
collaborators~\cite{denef} is relevant to understanding this issue.

Finally it would be very interesting to understand the role played by the enhan\c{c}on 
behind the horizon of our black holes. One might hope that there would be some sort of 
resolution of the time like singularity living there however as in  the case of the five
dimensional black holes~\cite{5d} this does not seem to be the case.  

The physics of the enhan\c{c}on is clearly a rich and very interesting 
topic for investigation. Above and beyond the applications to four dimensional
black holes presented in this paper and those 
mentioned briefly in the present discussion there are sure to be many interesting avenues
of pursuit.

\section*{Acknowledgements}
This research was supported by NSERC of Canada. 
I would especially
like to thank Clifford Johnson and Rob Myers for providing me with an early draft of
ref.~\cite{5d}, agreeing to proof read the present manuscript and sharing
numerous insights. I would also like to thank \O yvind Tafjord for useful 
conversations. Finally I would like to acknowledge the hospitality of the 
Physics Department at UCSB where the final stages of this research were carried out.

\end{document}